\documentclass[12pt]{iopart}
\usepackage{iopams}  
\usepackage[english]{babel}
\usepackage{graphicx}
\usepackage{cite}
\usepackage{bbm}

\begin{document}

\title{Entropy for the quantized field in the atom-field interaction: Initial mixed field}

\author{A. Z\'u\~niga-Segundo$^1$, R. Ju\'arez-Amaro$^2$, O. Aguilar-Loreto$^3$  and  H\'ector M. Moya-Cessa$^4$}
\address{$^1$
Instituto Polit\'ecnico Nacional. ESFM Departamento de F\'isica, Edificio 9 Unidad Profesional “Adolfo L\'opez Mateos, CP 07738 CDMX, Mexico\\
$^2$ Universidad Tecnol\'ogica de la Mixteca, Apdo. Postal 71,
69000 Huajuapan de Le\'on, Oax., Mexico\\
$^3$ Departamento de Ingenierias, CUCSur, Universidad de
Guadalajara CP
48900, Autl\'an de Navarro, Jal., Mexico \\
$^4$Instituto Nacional de Astrof\'{i}sica, \'{O}ptica y Electr\'{o}nica \\ Calle Luis Enrique Erro No. 1, Sta. Ma. Tonantzintla, Pue. CP 72840, Mexico }

%%%%%%%%%%%%%%%%%%%%%%%%%%%%%%%%%%%%%%%%%%%%%%%%%%%%%%%%%%%%%%%%%%%%%%%%
% Abstract
%%%%%%%%%%%%%%%%%%%%%%%%%%%%%%%%%%%%%%%%%%%%%%%%%%%%%%%%%%%%%%%%%%%
%
\begin{abstract}
We study the atom-field interaction when the field is in a mixture
of coherent states. We show that in this case it is possible to
calculate analytically the field entropy for times of the order of
twice the collapse time. Such analytical results are done with the
help of numerical analysis.
\end{abstract}
%%%%%%%%%%%%%%%%%%%%%%%%%%%%%%%%%%%%%%%%%%%%%%%%%%%%%%%%%%%%%%%%%%%%%%%%%%%%%%%%%%%%%%%%

\pacs{42.50.Ct, 42.50.-p, 42.50.Pq, 42.50.Dv}
%\ocis{ (050.5298) Photonic crystals;  (230.7370) Waveguides;  (350.5500) Propagation.}
%\submitto{\JPA}

\maketitle

%%%%%%%%%%%%%%%%%%%%%%%%%%%%%%%%%%%%%%%%%%%%%%%%%%%%%%%%%%%%%%%%%%%%%%%%%%%%%%%%%%%%%%%%

\section{Introduction}

It is well-known that the atomic inversion for a two-level atom interacting with a classical field \cite{Allen} undergoes Rabi oscillations \cite{Rabi0,Rabi}, which suffer collapses and revivals when instead of a classical field, a quantised field is considered. The shape of the revival of Rabi oscillations may be considered as an indicator of of the photon distribution of the  field being interacted with the two-level atom. For instance, when a squeezed field is considered, there exist so called ringing revivals of the Rabi oscillations that indicate such non-classical field \cite{Eberly,squeezed}. 

In the case of an initial coherent field, it is well known that its interaction with an initially excited two-level atom, produces, in phase space, the appearance of two counterrotating bumps that, after the so-called half revival time, $t_R/2$, produces approximately a superposition of coherent states \cite{Banacloche,Vidiela}.   Some (little) time before $t_R$, the bumps start to collide, which produces the revival of Rabi oscillations. If, instead of an initial coherent state, a superposition of two coherent states is considered, the revival of oscillations is divided by two, and the Rabi oscillations occur faster \cite{buzek}. However, if instead of the superposition of coherent states, a mixture of coherent states is initially considered, Rabi oscillations occur at $t_R$, just as in the case of a (single) coherent state. 

Although Rabi oscillations can give us some information about the initial state of the field, it is not enough, and we in general can not say, for instance, if the initial field was in a pure coherent state or in a mixture of coherent states, because both produce the same atomic inversion. One then can rely on the  entropy triangle inequality \cite{Araki}
\begin{eqnarray}
|S_A-S_B|\le S_{AB}\le S_A+S_B,
\end{eqnarray}
for two interacting subsystems, where $S_{AB}$ is the entropy of the total system and $S_A$ and
$S_B$ are the entropies of the subsystems A and B, respectively.
An important implication from this equation is that, if the total
entropy is zero, both subsystem's entropies will be equal.
Therefore, if in a lossless interaction, one can generate pure
initial states for subsystems A and B, their entropies will be
equal after evolution. For instance, in the (two-level) atom-field
interaction we may be able to find the entropy for the field given by
the entropy of the atom. A question arises: How to compute the
field's entropy in the case one of the subsystems, let us say the
quantised field, is not in a pure state? In this case the above
triangle inequality can not be applied and there is not a general
answer.

Phoenix and Knight have shown how the field entropy for an initial
coherent state can be calculated \cite{Phoenix1,Phoenix2} analytically. In
order to do so, the eigenstates and eigenvalues are obtained from the field density matrix. Moreover, Phoenix applied
later the same method to compute the entropy for a field subject
to decay \cite{Phoenix3}.   
Both calculations are rather complicated procedures even for a pure coherent
state. Although decay produces loss of information, it may be 
shown that information about an initial pure state can be obtained after decay happens \cite{Roversi1}. 

In this contribution we study a way to calculate the field entropy for an initial mixture of coherent states. We obtain analytical results with the help of numerical simulations.

\section{Entropies in the Jaynes-Cummings model}
 
Consider the Hamiltonian for the on-resonance interaction of an atom and a quantized field
\begin{eqnarray}
\hat{H}=\lambda(\hat{a}\hat{\sigma}_++\hat{a}^{\dagger}\hat{\sigma}_-)\;,
\end{eqnarray}
where for simplicity we do not take into account the free Hamiltonian terms (atom and field), and we only consider the interaction picture. The parameter $\lambda$ is the so-called Rabi frequency, the operators $\hat{a}$ and $\hat{a}^{\dagger}$ are the annihilation and creation operators, and the atomic operators, $\hat{\sigma}_-$ and $\hat{\sigma}_-$ are the lowering and raising Pauli matrices, respectively. If we consider an initial coherent state for the field, $|\alpha\rangle$ and an excited
state for the atom, we obtain the evolved (total) density matrix as
\begin{eqnarray}
\hat{\rho}= \left(
\begin{array}{ll}
|c\rangle\langle  c|  &|c\rangle\langle  s| \\
|s\rangle\langle  c|  & |s\rangle\langle s|
\end{array}
\right),
\end{eqnarray}
where the unnormalized wavefunctions are given by
\begin{eqnarray}
|c\rangle=\cos \lambda t\sqrt{\hat{a}\hat{a}^{\dagger}}|\alpha\rangle\;, \qquad
|s\rangle=-i\hat{V}^{\dagger}\sin \lambda t\sqrt{\hat{a}\hat{a}^{\dagger}}|\alpha\rangle\;,
\end{eqnarray}
with $\hat{V}=\frac{1}{\sqrt{\hat{a}\hat{a}^{\dagger}}}\hat{a}$ the London phase operator \cite{London,Susskind}.

We find the  field reduced density matrix as
\begin{eqnarray}
\hat{\rho}_F = |c\rangle\langle  c| + |s\rangle\langle  s|\;, \label{ccss}
\end{eqnarray}
while the atomic density matrix reads
\begin{eqnarray} \label{atomic}
\hat{\rho}_A =\left(
\begin{array}{ll}
\langle  c| c\rangle & \langle  s| c\rangle\\
\langle  c| s\rangle & \langle  s| s\rangle
\end{array}
\right)= \left(
\begin{array}{ll}
\rho_{11} & \rho_{12}\\
\rho_{21} & \rho_{22}
\end{array}
\right)\;.
\end{eqnarray}

Phoenix and Knight \cite{Phoenix1} looked for eigenstates and eigenvalues of $\rho_F$ in order to find the field entropy. Here we take a
different approach and will find first the field entropy operator, to finally obtaining the entropy.
 
\subsection{Relation between the powers of the reduced density matrices}

It may be easily proved that for two subsystems initially in pure states, after they interact, a relation between the powers of the
reduced density matrices may be obtained \cite{Moya}

\begin{eqnarray}\label{relacion}
\hat{\rho}_F^{n+1} =Tr_A\{\hat{\rho}\hat{\rho}_A^{n} \}\;, 
\end{eqnarray}
this helps to calculate functions of any of the density matrices, in particular the entropy operator $\hat{S}=-\ln \hat{\rho}$\;.

\subsection{Field entropy operator}

We will follow \cite{Moya} in order to obtain the field entropy operator. For this we first need to find $\hat{\rho}_A^{n}$, for this purpose we write

\begin{equation}
\hat{\rho}_A= \frac{{\bf 1}}{2}+\hat{R}\;,
\end{equation}
where
\begin{equation}
\hat{R}=\left(
\begin{array}{ll}
\frac{\delta}{2} & \rho_{12} \\
\rho_{21} & -\frac{\delta}{2}
\end{array}
\right)\;,
\end{equation}
with $\delta=\rho_{11}-\rho_{22}$ and ${\bf 1}$ is the $2\times 2$ unit density matrix. The powers of $\hat{\rho}_A$ are then  given by
\begin{equation}
\hat{\rho}_A^n=\left(\frac{{\bf 1}}{2}+\hat{R}\right)^n=
\sum_{m=0}^n \left(
\begin{array}{l}
n \\ m
\end{array}
\right) \frac{1}{2^{n-m}}\hat{R}^m\;.
\end{equation}
By splitting the above sum into two sums, one with odd powers of $\hat{R}$ and one with even powers and using
\begin{equation}
\hat{R}^{2m}=\epsilon^{2m}{\bf{1}}, \qquad
\hat{R}^{2m+1}=\frac{\hat{R}}{\epsilon}\epsilon^{2m+1}\;,
\end{equation}
with $\epsilon=\left(\frac{\delta^2}{4}+|\rho_{12}|^2\right)^{1/2}$ we can write

\begin{equation}
\hat{\rho}_A^n=\frac{{\bf 1}}{2}
\left[\left(\frac{1}{2}+\epsilon\right)^n+\left(
\frac{1}{2}-\epsilon\right)^n \right]+
\frac{\hat{R}}{2\epsilon}\left[\left(\frac{1}{2}+\epsilon\right)^n-\left(
\frac{1}{2}-\epsilon\right)^n \right]\;.
\end{equation}
We can finally write any power of $\rho_A$ in terms of itself and the unit matrix (as expected from Cayley-Hamilton's theorem (for $2\times 2$
matrices) \cite{Alle})
\begin{eqnarray}
 \hat{\rho}_A^n= G(n)\hat{\rho}_A
- || \hat{\rho}_A(t) || G(n-1) {\bf 1}\label{tothen}\;,
\end{eqnarray}
where
\begin{eqnarray}
G(n)=\frac{1}{2\epsilon}\left[\left(\frac{1}{2}+\epsilon\right)^n-\left(
\frac{1}{2}-\epsilon\right)^n\right]\;,
\end{eqnarray}
with the determinant $||\hat{\rho}_A(t)||=\frac{1}{4}-\epsilon^2$. Note that we have written  $\hat{\rho}_A^n$ in terms of $\hat{\rho}_A$ and the unity matrix. Alternatively one can express $\hat{\rho}_A^n$ in terms of the Chebyshev polynomials as

\begin{equation}
\hat{\rho}_A^n=||\hat{\rho}_A(t)||^{\frac{n}{2}}
\left[T_n\left(\frac{1}{2||\hat{\rho}_A(t)||^{\frac{1}{2}}}\right){\bf 1}+\frac{\hat{R}}{||\hat{\rho}_A(t)||^{\frac{1}{2}}}U_{n-1}\left(\frac{1}{2||\hat{\rho}_A(t)||^{\frac{1}{2}}}\right) \right],
\end{equation}
where
\begin{eqnarray}
\frac{1}{2\epsilon}\left[\left(\frac{1}{2}+\epsilon\right)^n-\left(
\frac{1}{2}-\epsilon\right)^n\right]&=&||\hat{\rho}_A(t)||^{\frac{n-1}{2}}U_{n-1}\left(\frac{1}{2||\hat{\rho}_A(t)||^{\frac{1}{2}}}\right)
\end{eqnarray}
and
\begin{eqnarray}
\frac{1}{2}\left[\left(\frac{1}{2}+\epsilon\right)^n+\left(
\frac{1}{2}-\epsilon\right)^n\right]&=&||\hat{\rho}_A(t)||^{\frac{n}{2}}T_{n}\left(\frac{1}{2||\hat{\rho}_A(t)||^{\frac{1}{2}}}\right)\;,
\end{eqnarray}
with $T_n$ and $U_n$ the  Chebyshev polynomials of the first and second kind respectively \cite{Arf}.

\begin{figure}[h!]
\begin{center}
\includegraphics[scale=0.5]{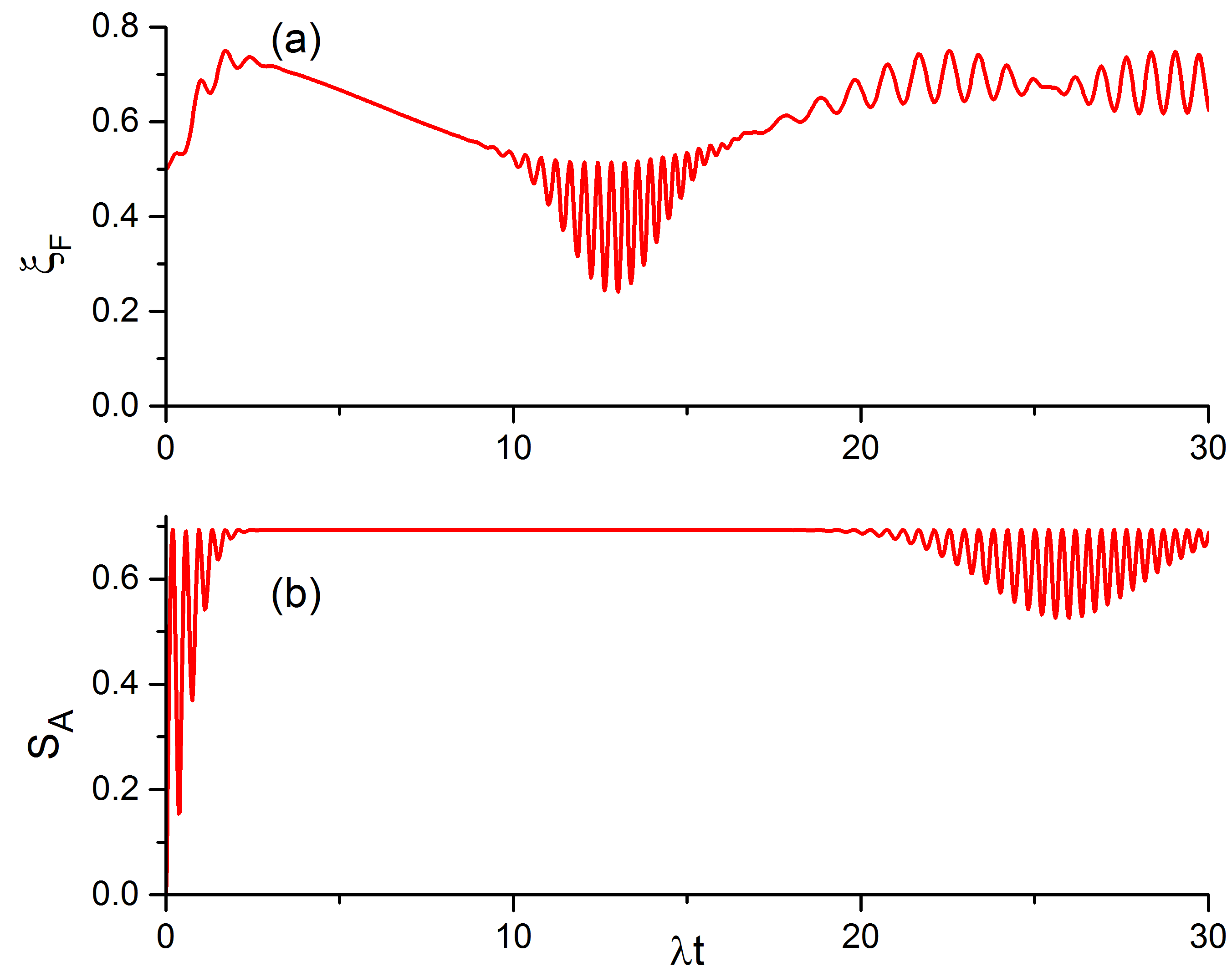}
\caption{(a) Field purity $\xi_F$ as a function of $\lambda t$ for the field initially prepared in statistical mixture of coherent states with $P_{\alpha}=P_{-\alpha}=\frac{1}{2}$, $\alpha=4.0$ and the atom in its excited state. (b) Atomic entropy $S_A$ against $\lambda t$ for the field initially prepared as in (a).}\label{fig1}
\end{center}
\end{figure}

We write the entropy operator as
\begin{eqnarray}
\hat{S}_A=\ln\hat{\rho}_A^{-1}=\ln({\bf 1}-\hat{\rho}_A)-\ln
||\hat{\rho}_A(t)||\;.
\end{eqnarray}
In the equation above we have used that $\hat{\rho}_A^2-\hat{\rho}_A+||\hat{\rho}_A(t)||{\bf 1}=0$, i.e. Cayley-Hamilton's theorem. We can 
expand in Taylor series $\ln(1-x)=-\sum_{n=1}^{\infty}\frac{x^n}{n}$ and use (\ref{tothen}) to find
\begin{eqnarray}
\hat{S}_A=F_1\hat{\rho}_A+F_2 {\bf 1}\;, \label{atent}
\end{eqnarray}
with
\begin{eqnarray}
\nonumber
& &F_1=\frac{1}{2\epsilon}\ln\left(\frac{1-2\epsilon}{1+2\epsilon}\right)\;,\\
& &F_2=-\frac{1}{2}\left[\ln||\hat{\rho}_A(t)||+
\frac{1}{2\epsilon}\ln\left(\frac{1-2\epsilon}{1+2\epsilon}\right)\right]\;.
\end{eqnarray}

\begin{figure}[h!]
\begin{center}
\includegraphics[scale=0.5]{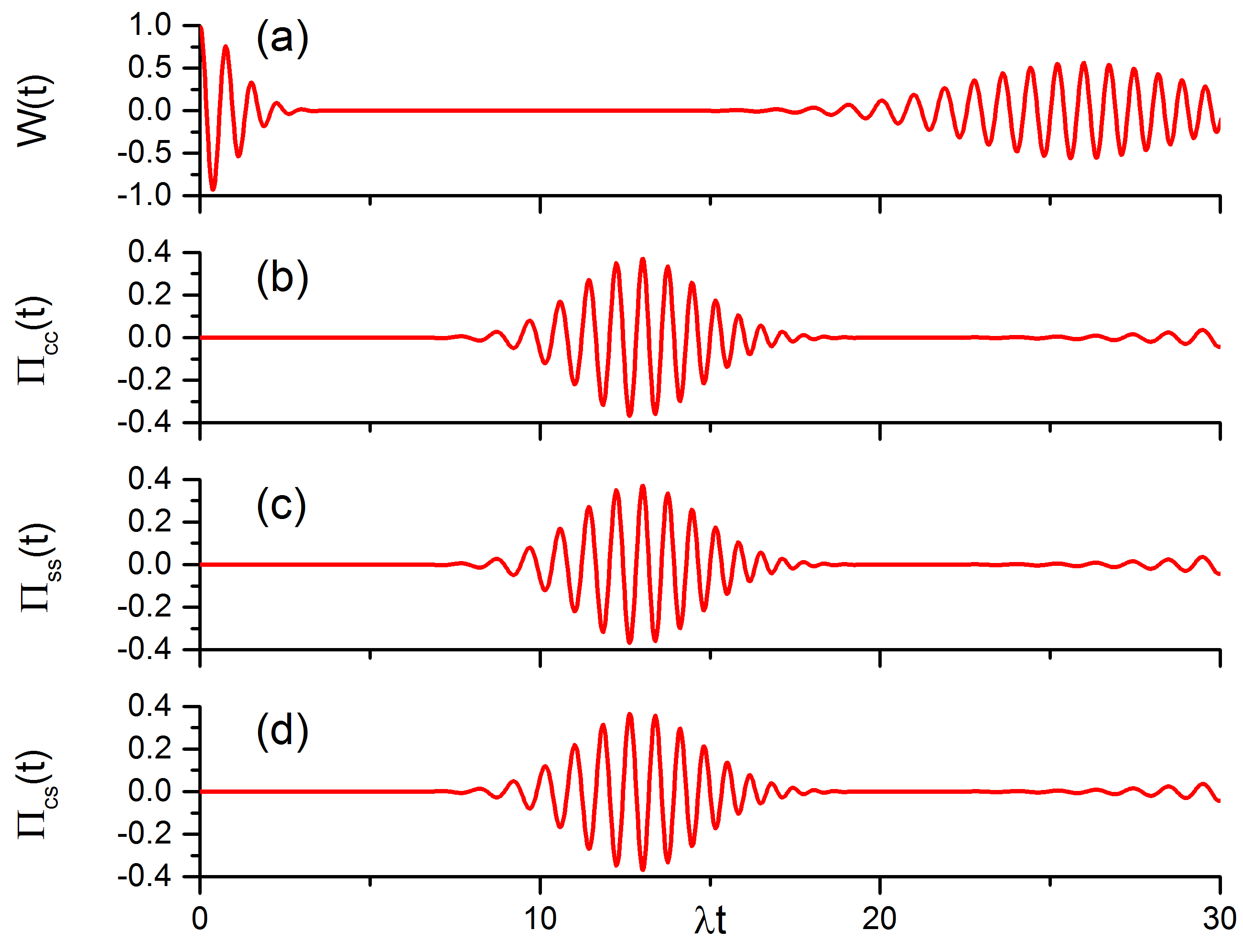}
\caption{(a) Time evolution of the total atomic inversion $W(t)=\langle{\hat{\sigma}}_z\rangle$ where the field and the atom were initially prepared as in Fig. (\ref{fig1}). Parity matrix elements of the states $|c\rangle$ and $|s\rangle$ (b)$\Pi_{cc}(t)$, (c)$\Pi_{ss}(t)$ and (d)$\Pi_{cs}(t)$; note they start in zero and continue being zero for a time longer than the collapse time.}\label{fig2}
\end{center}
\end{figure}

We now use the expression for $\hat{\rho}_B^n$ in terms of $\rho_A$ to write the field entropy operator in terms of the
atomic density operator
\begin{eqnarray}
\hat{S}_F=Tr_A\{\hat{\rho}(t)\hat{S}_A(t){\hat{\rho}}^{-1}_A(t)\}\;.
\label{ent-field}
\end{eqnarray}
By inserting (\ref{atent}) into (\ref{ent-field}) we obtain
\begin{eqnarray}
\hat{S}_F= Tr_A\{\hat{\rho}(t)(F_1 +F_2\hat{\rho}_A^{-1}(t))\}\;,
\end{eqnarray}
using the expression of the inverse of the atomic density operator in terms of the purity operator, the entropy may be written as
\begin{eqnarray}
\hat{S}_F= Tr_A\{\hat{\rho}(t)(F_1
+\frac{F_2}{||\hat{\rho}_A(t)||}[{\bf 1}-\hat{\rho}_A(t)])\}\;.
\label{29}
\end{eqnarray}
In terms of the field density matrix the field entropy operator is
\begin{eqnarray}
\hat{S}_F=\left(F_1+\frac{F_2}{||\hat{\rho}_A(t)||}\right)\hat{\rho}_F(t)
-\frac{F_2}{||\hat{\rho}_A(t)||}\hat{\rho}^2_F(t)\;, \label{S-f}
\end{eqnarray}
and from (\ref{ccss}) we can write $\hat{\rho}^2_F(t)$ as
\begin{equation}
\hat{\rho}^2_F=|c\rangle\langle c|\langle c|c\rangle
+|s\rangle\langle s| \langle s|s\rangle +|c\rangle\langle
s|\langle c|s\rangle +|s\rangle\langle c| \langle
s|c\rangle\label{density-f2}\;.
\end{equation}
The entropy operator for the field is written as
\begin{eqnarray}\label{S-f-f}
\hat{S}_F&=&\left(F_1+\frac{F_2\langle s|s\rangle}{||\hat{\rho}_A(t)||}\right)|c\rangle\langle c|+\left(F_1+\frac{F_2\langle c|c\rangle}{||\hat{\rho}_A(t)||}\right)|s\rangle\langle s|\\\nonumber
& &-\frac{F_2}{||\hat{\rho}_A(t)||}\left(\langle s|c\rangle |s\rangle\langle c|+\langle c|s\rangle|c\rangle\langle s|\frac{}{}\right).\nonumber
\end{eqnarray}

\section{Initial mixed field}

We now consider an initial field given by an statistical mixture of coherent states,
$\hat{\rho}_F^{mix}(0)=P_{\alpha}|\alpha\rangle\langle\alpha|+P_{-\alpha}|-\alpha\rangle\langle-\alpha|$, and the atom in its excited state. The quantities  $P_{\alpha}$ and $P_{-\alpha}$ are the probabilities that the quantized field is either in the coherent state with amplitude ${\alpha}$ or $-{\alpha}$, respectively, with $P_{\alpha}+P_{-\alpha}=1$. The evolved density matrix for such initial conditions is

\begin{eqnarray}
\hat{\rho}^{mix} = \left(
\begin{array}{ll}
P_{\alpha}|c\rangle\langle  c| +P_{-\alpha} \Pi|c\rangle\langle  c|\Pi &P_{\alpha}|c\rangle\langle  s| - P_{-\alpha}\Pi|c\rangle\langle  s|\Pi\\
P_{\alpha}|s\rangle\langle  c| - P_{-\alpha}\Pi|s\rangle\langle  c|\Pi & P_{\alpha}|s\rangle\langle  s| + P_{-\alpha}\Pi|s\rangle\langle  s|\Pi\;,
\end{array}
\right)
\end{eqnarray}
where $\Pi=(-1)^{\hat{a}^{\dagger }\hat{a}}$ is the so-called parity operator, with the relation  $\Pi|\alpha\rangle=|-\alpha\rangle$. It is clear that, by tracing over the field basis we obtain the atomic density matrix

\begin{eqnarray}
\hat{\rho}_A^{mix} =\left(
\begin{array}{ll}
\langle  c| c\rangle & (P_{\alpha}-P_{-\alpha})\langle  c| s\rangle\\
(P_{\alpha}-P_{-\alpha})\langle  s| c\rangle& \langle  s| s\rangle
\end{array}
\right)\;,
\end{eqnarray}
and by tracing over  the atomic basis we obtain the field density matrix

\begin{eqnarray}
\hat{\rho}_F^{mix} = P_{\alpha}(|c\rangle\langle  c|+|s\rangle\langle  s|) +P_{-\alpha}\Pi (|c\rangle\langle  c|+  |s\rangle\langle  s|)\Pi\;,
\end{eqnarray}
and from it we can calculate

\begin{eqnarray} \label{27}\nonumber
(\hat{\rho}_F^{mix})^2 &=&I_{cc}(P_{\alpha}^2|c\rangle\langle  c| +P_{-\alpha}^2 \Pi|c\rangle  \langle c| \Pi) +P_{\alpha}P_{-\alpha}(\Pi|c\rangle\langle  c| +|c\rangle\langle  c|\Pi)\Pi_{cc}\\ \nonumber
&+&I_{ss}(P_{\alpha}^2|s\rangle\langle  s| +P_{-\alpha}^2 \Pi|s\rangle\langle  s|\Pi) +P_{\alpha}P_{-\alpha}(\Pi|s\rangle\langle  s|+|s\rangle\langle  s|\Pi)\Pi_{ss}\\\nonumber
&+&I_{cs}(P_{\alpha}^2|c\rangle  \langle s|+ P_{-\alpha}^2\Pi|c\rangle  \langle s|\Pi) +P_{\alpha}P_{-\alpha}(\Pi|c\rangle\langle  s| +|c\rangle\langle  s|\Pi)\Pi_{cs}\\
&+&I_{sc}(P_{\alpha}^2|s\rangle\langle  c| + P_{-\alpha}^2\Pi|s\rangle\langle  c|\Pi) +P_{\alpha}P_{-\alpha}(\Pi|s\rangle\langle  c| +|s\rangle\langle  c|\Pi)\Pi_{sc}\;,
\end{eqnarray}
in order to have an idea of the behaviour of the field entropy we have calculated the purity parameter $\xi_F=1-Tr{(\hat{\rho}_F^{mix})^2}$ and the atomic entropy $S_A$ as it is shown in Fig. \ref{fig1}(a) and (b) respectively. In the above equation we have
defined $I_{jk},=\langle  j| k\rangle$ and $\Pi_{jk}=\langle  j| \Pi|k\rangle$, $j,k=c,s$. In Fig. \ref{fig2}(a) we have plotted time evolution of the total atomic inversion $W(t)=\langle{\hat{\sigma}}_z\rangle$ and the figures \ref{fig2}(b-c) the $\Pi_{cc}$, $\Pi_{ss}$ and $\Pi_{cs}$ respectively. 

Note that they start in zero and continue being zero for a time longer than the collapse time, in the figure for times up to $\lambda t\approx 8$. For such periods of time we may disregard the second term in each of the rows of equation (\ref{27}), yielding exactly

\begin{eqnarray} \label{28}\nonumber
(\hat{\rho}_F^{mix})^2 &= &P_{\alpha}^2\left(I_{cc}|c\rangle\langle  c| +I_{ss}|s\rangle\langle  s| +I_{cs}|c\rangle  \langle s|+I_{sc}|s\rangle\langle  c|\right) \\
&+&P_{-\alpha}^2\Pi\left(I_{cc}|c\rangle\langle  c| +I_{ss}|s\rangle\langle  s| +I_{cs}|c\rangle  \langle s|+I_{sc}|s\rangle\langle  c|\right)
\Pi\;.
\end{eqnarray}
Equation (\ref{28}) tells us that, for the periods in which $\Pi_{jk}=0$ we may separate the calculation of the entropy for the field, obtaining terms related only to the state $|\alpha\rangle$ or  $|-\alpha\rangle$. Therefore, we can apply the method of Section 2 to each of them. Before that, let us rewrite a generalization of equation (\ref{28}) 

\begin{eqnarray} 
(\hat{\rho}_F^{mix})^k = P_{\alpha}^k\left(|c\rangle\langle  c|+|s\rangle\langle  s|\right)^k + P_{-\alpha}^k\Pi\left(|c\rangle\langle  c|+|s\rangle\langle  s|\right)^k
\Pi\;,
\end{eqnarray}
and therefore, a function of it may be written as
\begin{eqnarray} 
f(\hat{\rho}_F^{mix}) = f\left[P_{\alpha}\left(|c\rangle\langle  c|+|s\rangle\langle  s|\right)\right] +\Pi f\left[ P_{-\alpha}\left(|c\rangle\langle  c|+|s\rangle\langle  s|\right)\right]
\Pi\;,
\end{eqnarray}
so that
\begin{eqnarray} \nonumber
\hat{S}_F^{mix}=-\ln(\hat{\rho}_F^{mix}) = &-&\ln\left[\left(|c\rangle\langle  c|+|s\rangle\langle  s|\right)\right] -\Pi \ln\left[ \left(|c\rangle\langle  c|+|s\rangle\langle  s|\right)\right]
\Pi \\&-&\ln P_{\alpha}-\ln P_{-\alpha}\;,
\end{eqnarray}
or
\begin{eqnarray} 
\hat{S}_F^{mix}= \hat{S}_F+\Pi \hat{S}_F\Pi -\ln P_{\alpha} P_{-\alpha}\;,
\end{eqnarray}
where $\hat{S}_F$ is given in (23). Finally we can obtain the entropy by tracing the entropy operator times the density matrix
\begin{eqnarray}
{S}_F^{mix}= Tr_F\{\hat{S}_F^{mix}\hat{\rho}_F^{mix}\}=S_F-P_{\alpha}\ln P_{\alpha}-P_{-\alpha}\ln P_{-\alpha}\;,
\end{eqnarray}
where $S_F=Tr\{\hat{S}_F\hat{\rho}_{F}\}.$ In Figure 3 we plot the entropy as a function of $\lambda t$. It may be noticed in the above equation that the entropy for the entropy for the mixture of coherent states splits into two parts, one given by the entropy of a single coherent states and the other one given just by the mixture probabilities. This figure should be compared with Fig. 1(a) which has oscillations about the region of $\lambda t \approx 13$ reflecting the term we neglected in equation (34).

\begin{figure}[h!]
\begin{center}
\includegraphics[scale=0.5]{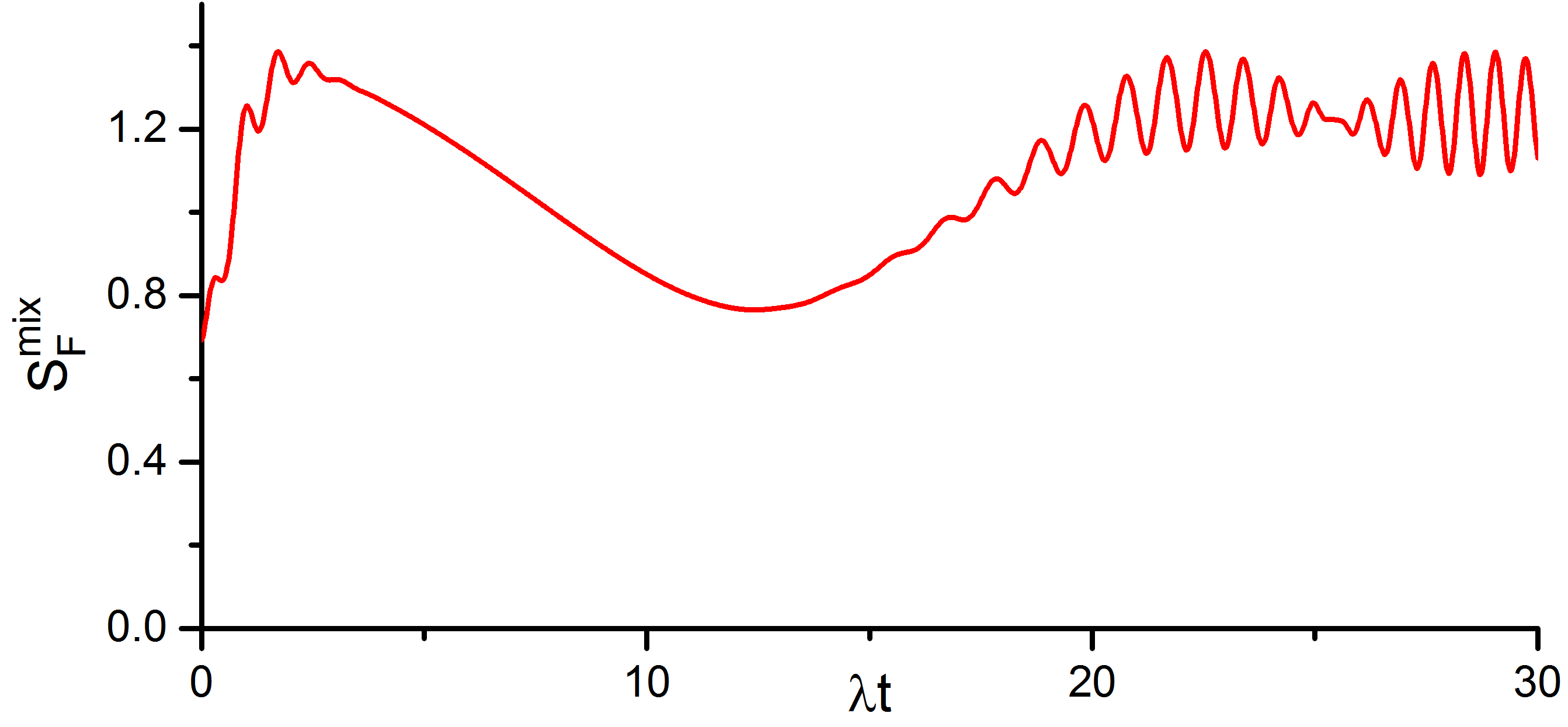}
\caption{Time evolution of the field mixed entropy $S_F^{mix}$ where the field initially was prepared in statistical mixture of coherent states with $P_{\alpha}=P_{-\alpha}=\frac{1}{2}$, $\alpha=4.0$ and the atom was in its excited state as in Fig. (\ref{fig1}).}\label{fig3}
\end{center}
\end{figure}

\section{Conclusions}
By noticing that the quantities $\Pi_{jk}=\langle  j| \Pi|k\rangle$, with $j,k=c,s$, are zero for relevant periods of time, for such periods we can calculate, { \it exactly} the entropy for an initial field composed by a statistical mixture of coherent states. We could argue that because of the {\it cuasi-orthonormality} of different coherent states with (large) different amplitudes it is possible to calculate the entropy in an analytical form, but this is not the case: We are truly relaying on the numerical properties of $\Pi_{jk}$, i.e., in the fact that, for large periods of time, these quantities are zero. 

\end{document}